\documentclass[10pt]{iopart}
\usepackage{psfig}  

\begin{document}

%
\jl{6}

\title[Errors on the inverse problem\ldots]{Errors on the inverse problem
       solution for a noisy spherical gravitational wave antenna}

\author{Stephen M Merkowitz\dag, J Alberto Lobo\ddag\footnote[3]{To whom
	correspondence should be addressed.} and M Angeles Serrano\ddag}

\address{\dag\ Nuclear Physics Laboratory, University of Washington,
	 Seattle WA 98195-4290, USA, and INFN Laboratori Nazionali di
	 Frascati, Via Enrico Fermi 40, I-00044 Frascati (Roma), Italy}

\address{\ddag\ Departament de F\'\i sica Fonamental,
	 Universitat de Barcelona, Diagonal 647, 08028 Barcelona, Spain.}

\begin{abstract}
A single spherical antenna is capable of measuring the direction and
polarization of a gravitational wave. It is possible to solve the inverse
problem using only linear algebra even in the presence of noise. The
simplicity of this solution enables one to explore the error on the
solution using standard techniques. In this paper we derive the error on the
direction and polarization measurements of a gravitational wave. We show that
the solid angle error and the uncertainty on the wave amplitude are direction
independent. We also discuss the possibility of determining the
polarization amplitudes with isotropic sensitivity for any given gravitational 
wave source.
\end{abstract}


\submitted


\section{Introduction}

A spherical gravitational wave antenna ideally has equal sensitivity to
gravitational waves from all directions and polarizations and is able to
determine the directional information and tensorial character of a
gravitational wave.  The solution for the inverse problem for a noiseless
antenna has been known for some time \cite{Wagoner_Pavia_1976}, and an 
analytic solution for an noisy antenna was recently found 
\cite{Merkowitz_PRD_1998}.  These solutions are quite elegant as they only 
require linear algebra to estimate the wave direction and polarization from 
the detector outputs.

By monitoring the five quadrupole modes of an elastic sphere, one has a 
direct measurement of the effective force of a gravitational wave on the
sphere \cite{Lobo_PRD_1995}. The standard technique for doing so on
resonant detectors is to position resonant transducers on the surface of
the sphere that strongly couple to the quadrupole modes.  A number of
proposals have been made for the type and positions of the transducers
\cite{Johnson_PRL_1993,Lobo_EPL_1996,Zhou_PRD_1995}.  What all of these 
proposals have in common is that the outputs of the transducers are 
combined into five ``mode channels'' $g_m(t)$ that are constructed to have 
a one-to-one correspondence with the quadrupole modes of the sphere and 
thus the spherical amplitudes of the gravitational
wave~\cite{Merkowitz_PRD_1995,Lobo_CQG_1998}.

The mode channels $g_m\/$ can be collected to form a ``detector response'' 
matrix

\begin{equation}
   \bi{A} =
   \left[\begin{array}{ccc}
      g_{1} - \frac{1}{\sqrt{3}}g_{5} & g_{2} &  g_{4} \\
      g_{2} &  -g_{1} - \frac{1}{\sqrt{3}}g_{5} &  g_{3} \\
      g_{4} &  g_{3} &  \frac{2}{\sqrt{3}} g_{5}
   \end{array} \right],
   \label{eqn:cartesian_strain_tensor}
\end{equation}
which, in the absence of noise in the detector, is equal to the GW strain
tensor, expressed in lab frame coordinates. The latter tensor has the
canonical form

\begin{equation}
   \bi{H} = \left[ 
   \begin{array}{ccc}
      h_+      & h_\times & 0 \\
      h_\times &  -h_+    & 0 \\
      0            & 0            & 0
   \end{array} \right],
   \label{eqn:wave_strain}
\end{equation}
in the wave frame, and is related to $\bi{A}$ by an orthogonal
transformation ---a rotation. $\bi{H}$ clearly has an eigenvector,
$\bi{v}_3$, say, with zero eigenvalue which corresponds to the wave
propagation direction. The same therefore applies to $\bi{A}$, and this
enables the determination of the wave direction by a straightforward algebraic
procedure directly from detector data: it is the eigenvector of $\bi{A}$
with null eigenvalue.

Things change when the detector is noisy: noise gets added to
the signal in the mode channels, destroying the equivalence between
the data matrix $\bi{A}$ and the signal matrix $\bi{H}$. However, it
has been shown that under ideal conditions of the noise a modified version
of the above procedure can be used \cite{Merkowitz_PRD_1998}: the eigenvector
of the noisy $\bi{A}$ with eigenvalue {\it closest to zero\/} is the
{\it best\/} approximation to the actual incidence direction of the
gravitational wave.

In this paper we shall be taking an {\it analytic\/} approach to the
diagonalization problem, whereby errors in the estimated GW parameters can
be assessed to any desired degree of accuracy. Inherent in this approach is
the unambiguous definition of the incidence direction estimate, as well as
a Cartesian coordinate convention for it, which rids us of the ambiguities
intrinsically associated to the Euler angle characterization for incidence
directions near the Poles. Errors in these quantities will be shown to be
incidence direction independent. In addition, we shall also address the
problem of estimating the GW amplitudes $h_+$ and $h_\times$, and their
errors. Previous authors \cite{Zhou_PRD_1995,Stevenson_PRD_1997} found it
impossible to give isotropic estimates of these quantities, a very strange
result for a spherical detector.  We explain why these results come about
and we show how the problem can be solved by properly including all
the necessary information.

The paper is organized as follows. We begin in section~\ref{sec:A_ev} by
deriving analytic expressions for the eigenvalues of $\bi{A}$. We then
use these expressions to find the first order statistical errors in the
eigenvalues in section~\ref{sec:uncertainties_ev}, followed by the direction
estimation error in section~\ref{sec:dir_error}. Higher order corrections to
these errors are presented in section~\ref{sec:corrections}.
In section~\ref{sec:polarization} we discuss the errors on the polarization
amplitude estimates. We explain why past solutions
have direction dependent errors on these quantities and we describe a
maximum likelihood algorithm, based a hypothesis on the physical nature
of the source, that fulfills the natural property of source location
independence.

\section{Detector response eigenvalues}
\label{sec:A_ev}

The detector response matrix $\bi{A}$ is symmetric and traceless and has 
the eigenvalue equation

\begin{equation}
	\bi{A}\bi{v}_k = \lambda_k\bi{v}_k, \; k=1,2,3,
	\label{eqn:noisy_ev}
\end{equation}
where the eigenvectors are normalized in the usual way

\begin{equation}
	\bi{v}_k \cdot \bi{v}_l = \delta_{kl}.
\end{equation}
Expanding equation \eref{eqn:noisy_ev} we find

\begin{equation}
	\lambda_k^3 - g^2 \lambda_k - D = 0,
\end{equation}
where we have defined

\numparts
\begin{eqnarray}
	g^2 \equiv g_1^2 + g_2^2 + g_3^2 + g_4^2 + g_5^2 \\[1 ex]
	D \equiv \det(\bi{A}).
\end{eqnarray}
\endnumparts
Solving this cubic equation we find the eigenvalues of $\bi{A}$ to be

\begin{equation}
        \lambda_k = -\frac{2}{\sqrt{3}}\;g\,\cos\theta_k, \ k=1,2,3,
        \label{eq:roots}
\end{equation}
where

\begin{equation}
	\theta_k = \frac{\theta + 2(k-1)\pi}{3},\ \mbox{and}\ 
        \cos\theta = -\frac{3\sqrt{3}}{2}\;\frac{D}{g^3}.
        \label{eq:1.9}
\end{equation}

The $k=3$ eigenvalue is identically zero in the absence of noise, so it
will generally be the one closest to zero in the presence of noise. Random
fluctuations may eventually change this (more likely for low SNR), but we
shall always take the corresponding eigenvector $\bi{v}_3$ as the best
approximation to the direction of the source.
The amplitude of the wave $h$ can be calculated in many ways from the mode 
channels (for example, $g\/$ is an estimate for the amplitude), but the 
semi-difference of the other two eigenvalues will 
give the best estimate \cite{Merkowitz_PRD_1998},
\begin{equation}
	h = \frac{1}{2}\,\left(\lambda_2 - \lambda_1\right).     
	\label{eq:estim_h}
\end{equation}

\section{Eigenvalue errors}
\label{sec:uncertainties_ev}

We assume that the mode channels $g_m$ have uncorrelated noise with zero
mean and variance $\sigma_{g_m}^2 \equiv E\left\{(\delta g_m)^2\right\}$.
The lowest order statistical errors in the eigenvalues are easily 
calculated by
\begin{equation}
	\sigma_{\lambda_k}^2
	= 
	\sum_{m=1}^5
	\left(\frac{\partial\lambda_k}{\partial g_m}\right)^2
	\sigma_{g_m}^2,
	\label{eq:ev_error}
\end{equation}
where the derivatives of the eigenvalues are given by
\begin{equation}
	\frac{\partial\lambda_k}{\partial g_m} = 
        -\frac{2}{\sqrt{3}}\;\frac{g_m}{g}\,\cos\theta_k +
        \frac{\partial D}{\partial g_m}\,\sin\theta_k,
        \label{eq:derl}
\end{equation}
and the derivatives of the determinant $D\/$ are

\numparts
\begin{eqnarray}
  \frac{\partial D}{\partial g_1} 
  & = &
  -\frac{4}{\sqrt{3}}g_1g_5 - g_3^2 + g_4^2,
  \\
  \frac{\partial D}{\partial g_2} 
  & = &
  -\frac{4}{\sqrt{3}}g_2g_5 + 2g_3g_4,
  \\
  \frac{\partial D}{\partial g_3} 
  & = &
  -2g_3\left(g_1-\frac{g_5}{\sqrt{3}}\right) + 2g_2g_4,
  \\
  \frac{\partial D}{\partial g_4}
  & = &
  2g_4\left(g_1+\frac{g_5}{\sqrt{3}}\right) + 2g_2g_3,
  \\
  \frac{\partial D}{\partial g_5} 
  & = &
  \frac{1}{\sqrt{3}}\,\left(-2g_1^2-2g_2^2+g_3^2+g_4^2+2g_5^2\right).
\end{eqnarray}
\endnumparts
If the variances on the mode channels are equal then it is easily seen that
equations \eref{eq:ev_error} and \eref{eq:derl} lead to

\begin{equation}
	\sigma_{\lambda_k}^2 = \frac{4}{3}\,\sigma_g^2,\ k=1,2,3,
        \label{eq:iso}
\end{equation}
where $\sigma_g^2$ is the variance on any one mode channel. Note that from
equation \eref{eq:iso} all three eigenvalues have equal variance to first
order. Cross correlations between these eigenvalues are also easily
calculated, and for equal mode channel variances they are equal for all
pairs $(\lambda_k,\lambda_k')$:

\begin{equation}
	\sigma_{\lambda_k\lambda_k'} = -\frac{2}{3}\,\sigma_g^2,\ k,k'=1,2,3.
	\label{eq:isocross}
\end{equation}

\begin{figure}
\psfig{file=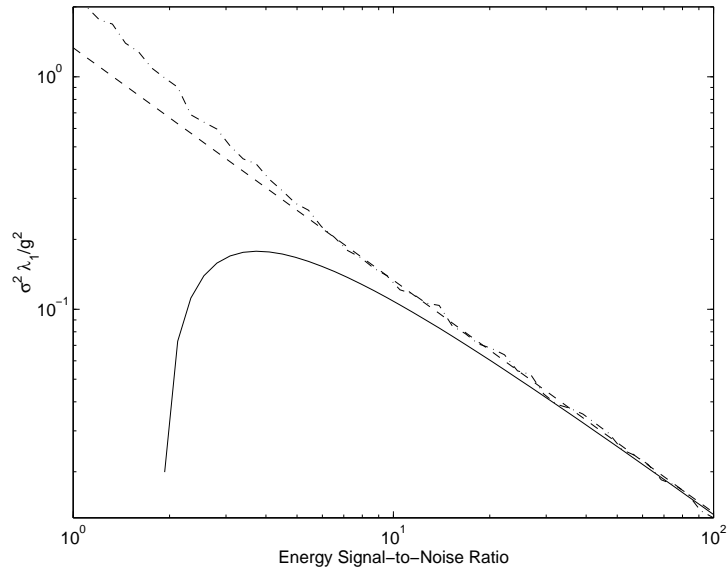,height=15cm,rheight=7.8cm,bbllx=-3cm,bblly=-6.4cm,bburx=14.2cm,bbury=21cm}

\caption{The results of a numerical simulation describing the variance of
the first eigenvalue for a range of SNR.  The dot-dashed line was computed
by a 1000 trial Monte Carlo simulation for the range of SNR.
The dashed line is the error found from the first order analytic
expression and the solid line includes the second order corrections.}

\label{fig:ev_error_1}
\end{figure}

\begin{figure}
\psfig{file=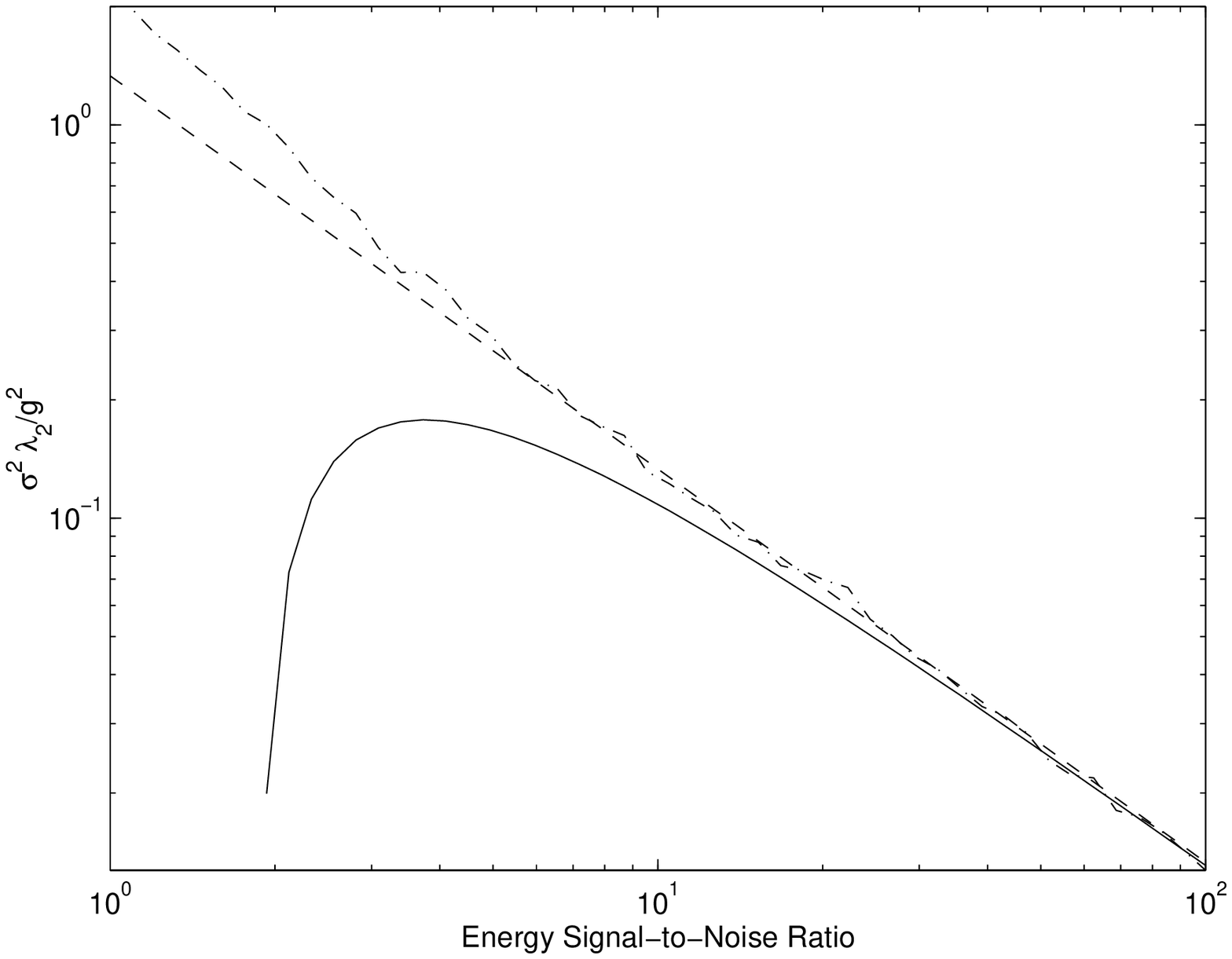,height=15cm,rheight=7.8cm,bbllx=-3cm,bblly=-6.4cm,bburx=14.2cm,bbury=21cm}

\caption{The results of a numerical simulation describing the variance of
the second eigenvalue for a range of SNR.  The dot-dashed line was computed
by a 1000 trial Monte Carlo simulation for the range of SNR.
The dashed line is the error found from the first order analytic
expression and the solid line includes the second order corrections.}

\label{fig:ev_error_2}
\end{figure}

\begin{figure}
\psfig{file=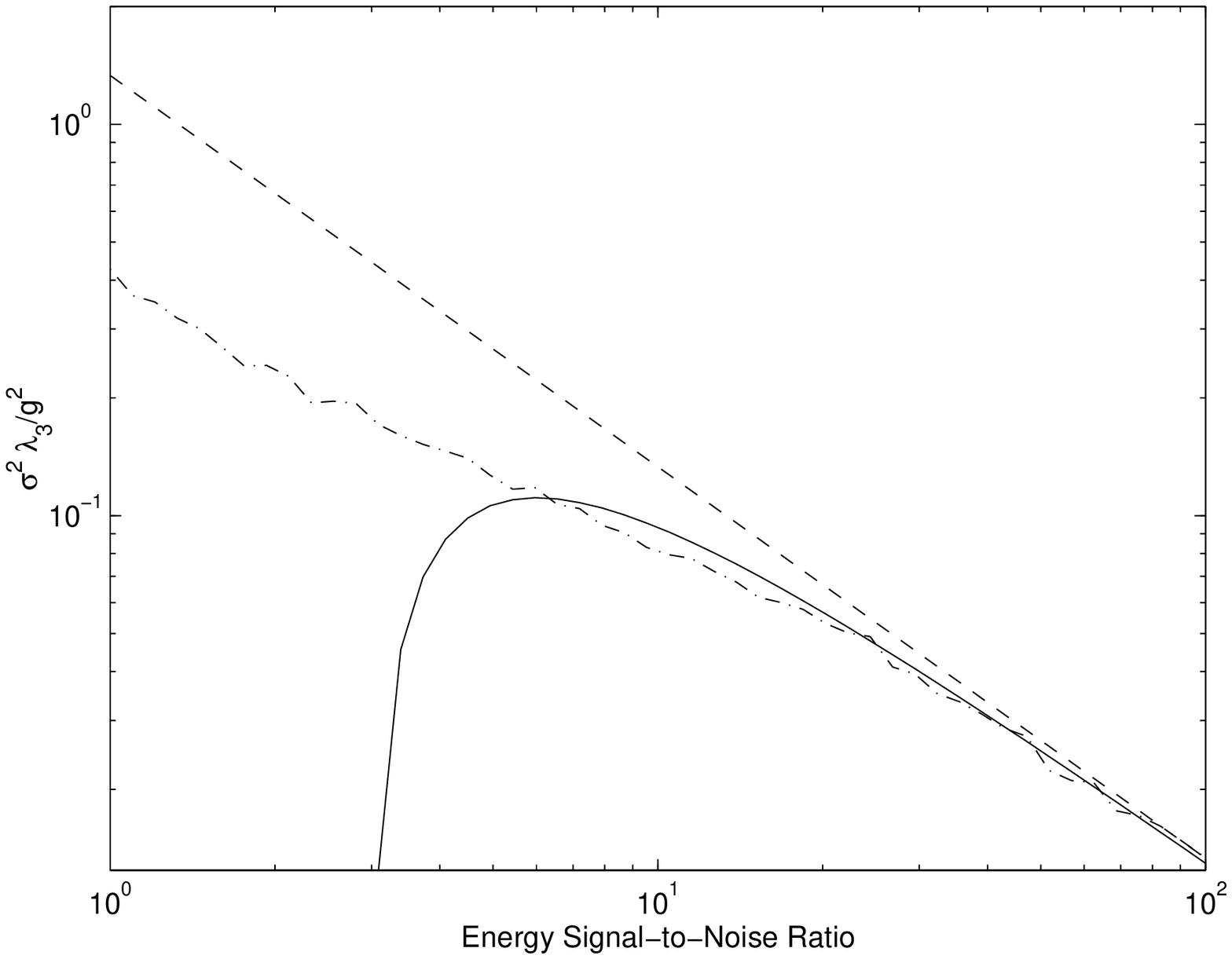,height=15cm,rheight=7.8cm,bbllx=-3cm,bblly=-6.4cm,bburx=14.2cm,bbury=21cm}

\caption{The results of a numerical simulation describing the variance of
the third eigenvalue for a range of SNR.  The dot-dashed line was computed
by a 1000 trial Monte Carlo simulation for the range of SNR.
The dashed line is the error found from the first order analytic
expression and the solid line includes the second order corrections.}

\label{fig:ev_error_3}
\end{figure}

Shown in Figs.~\ref{fig:ev_error_1}-\ref{fig:ev_error_3} is this variance
as a function of the ${\rm  SNR} = g^2/\sigma_g^2$.  Also shown is the
results of a Monte Carlo type simulation of the errors which take into
account the higher order perturbations at low values of SNR.  As shown, the
analytic expressions match the simulated errors for high SNR.  For low SNR
discrepancies arise between the analytic and simulated values.  This is
within expectation since equation \eref{eq:iso} is only accurate for large
SNR. Higher order correction will be considered below.

\section{Direction estimation error}
\label{sec:dir_error}

We assume that the eigenvector $\bi{v}_3$ points in the propagation 
direction of the gravitational wave.  We want to estimate the fluctuations 
in the determination of this direction caused by the presence of noisy 
fluctuations in the mode channels.  We represent with $\delta$ a difference 
between a given quantity and its ideal value if there were no noise (i.e.  
$\delta\bi{v}_3$ is the difference between the position calculated from 
noisy data and its real position in the sky).  We now take equation
\eref{eqn:noisy_ev} for $k=3$ and consider fluctuations in it.  If 
these are not too large (high SNR) we can retain only first order terms
\begin{equation}
	\left[\bi{A}-\lambda_3\right]\delta\bi{v}_3 
	=
	-\left[\delta\bi{A}-\delta\lambda_3\right]\bi{v}_3.
	\label{pert}  
\end{equation}
This is an equation for $\delta\bi{v}_3$, but the matrix 
$\left[\bi{A}-\lambda_3\right]$ is not invertible.  The only consequence 
of this is that we cannot determine the component of $\delta\bi{v}_3$ 
which is parallel to $\bi{v}_3$ itself.  The orthogonal components (those 
parallel to $\bi{v}_1$ and $\bi{v}_2$) can easily be found by 
multiplying equation \ref{pert} on the left by $\bi{v}_1$ and $\bi{v}_2$
\begin{eqnarray}
	\bi{v}_1\cdot\delta\bi{v}_3 
	& = &
	-\frac{1}{\lambda_1-\lambda_3}\;\bi{v}_1\delta\bi{A}\bi{v}_3,
	\\
	\bi{v}_2\cdot\delta\bi{v}_3
	& = &
	-\frac{1}{\lambda_2-\lambda_3}\;\bi{v}_2\delta\bi{A}\bi{v}_3.
\end{eqnarray}

An appropriate assessment of the error on a direction measurement is the 
solid angle error $\Delta\Omega$.  Since $|\bi{v}_3|=1$, this error is 
given by
\begin{equation}
  \Delta\Omega = \pi\left|\Delta\bi{v}_3\right|^2.
  \label{eq:omega}
\end{equation}
where $|\Delta\bi{v}_3|^2$ is the quadratic error in the
determination of $\bi{v}_3$. To find it we need to calculate the
expectation of the squared modulus of the above fluctuations,

\begin{equation}
	\fl\left|\Delta\bi{v}_3\right|^2 = E\left\{
		(\bi{v}_1\cdot\delta\bi{v}_3)^2 
		+ (\bi{v}_2\cdot\delta\bi{v}_3)^2\right\} =
	E\left\{\left|
		\frac{\bi{v}_1\delta\bi{A}\bi{v}_3}{\lambda_1-\lambda_3}
		\right|^2 + \left|
		\frac{\bi{v}_2\delta\bi{A}\bi{v}_3}{\lambda_2-\lambda_3}
		\right|^2\right\}.
  \label{spqr}
\end{equation}
First order calculations only require us to take expectations in 
$\delta\bi{A}$, while leaving the rest untouched,

\begin{equation}
  \delta\bi{A} = \sum_{m=1}^5 \bi{A}_m\delta g_m,
\end{equation}
where we have defined

\begin{equation}
   \bi{A}_m\equiv\frac{\partial \bi{A}}{\partial g_m},
\end{equation}
Explicitly,

\numparts
\begin{eqnarray}
	\fl\qquad\ \ \ \ \bi{A}_1 
	= 
	\left[\begin{array}{ccc}
	1 &  0 & 0 \\
	0 & -1 & 0 \\
	0 &  0 & 0
	\end{array}\right],
	\ \bi{A}_2 
	=
	\left[\begin{array}{ccc} 
	0 & 1 & 0 \\
	1 & 0 & 0 \\
	0 & 0 & 0
	\end{array}\right],
	\ \bi{A}_3 
	=
	\left[\begin{array}{ccc}
	0 & 0 & 0 \\
	0 & 0 & 1 \\
	0 & 1 & 0
	\end{array}\right], \\[1 em]
	\bi{A}_4
	= 
	\left[\begin{array}{ccc}
	0 & 0 & 1 \\
	0 & 0 & 0 \\
	1 & 0 & 0 
	\end{array}\right],
	\ \bi{A}_5 
	=
	\left[\begin{array}{ccc} 
	-\frac{1}{\sqrt{3}} & 0 & 0 \\
	0 & -\frac{1}{\sqrt{3}} & 0 \\
	0 & 0 & \frac{2}{\sqrt{3}}
	\end{array}\right].
\end{eqnarray}
\endnumparts
We thus have

\begin{equation}
	|\Delta\bi{v}_3|^2 =
	\sum_{m=1}^5\left[\left(
	\bi{v}_1\bi{A}_m\bi{v}_3
	\right)^2
	+ \left(
	\bi{v}_2\bi{A}_m\bi{v}_3
	\right)^{2}
	\right]\,\frac{\sigma_{g_m}^2}{g^2}.
	\label{eq:deltav}
\end{equation}

Again, setting the variances on the mode channels equal to $\sigma_g^2$,
the sum in equation \eref{eq:deltav} can easily be done, giving
\begin{equation}
	|\Delta\bi{v}_3|^2 = 2\,\frac{\sigma_g^2}{g^2}.
	\label{eq:dv3}
\end{equation}

From equation \eref{eq:dv3} we see that the error in the incidence direction
is independent of this direction, as expected of an omnidirectional antenna.
Substituting this into \eref{eq:omega} we find

\begin{equation}
        \Delta\Omega = \frac{2\pi}{SNR}.
	\label{eq:soe}
\end{equation}

\begin{figure}
\psfig{file=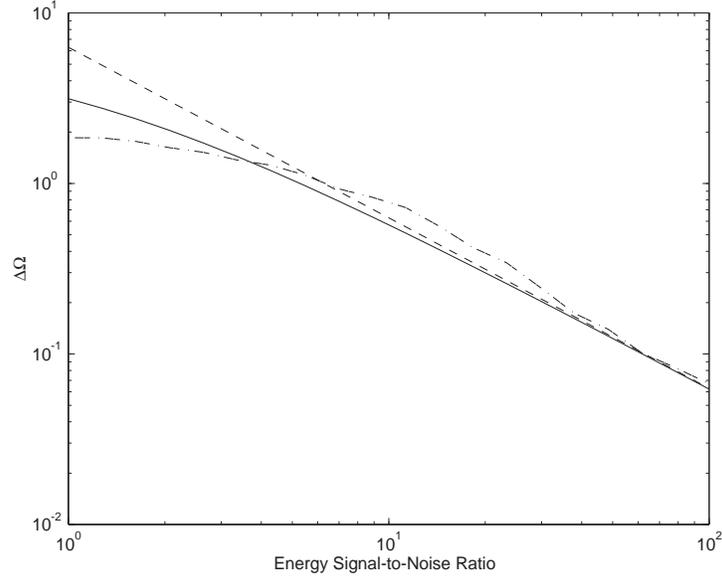,height=15cm,rheight=7.8cm,bbllx=-3cm,bblly=-6cm,bburx=14.2cm,bbury=21.4cm}

\caption{The results of a numerical simulation describing the solid angle
direction estimation error $\Delta\Omega$ of a source direction measurement
due to a finite signal-to-noise ratio.  The dot-dashed line was computed by
a 1000 trial Monte Carlo simulation for the range of SNR.  The dashed line
is the error found from the first order analytic expression and the solid
line includes the second order corrections.}

\label{fig:solid_angle_error}
\end{figure}

This expression is in perfect agreement with the solid angle estimation 
error found by Zhou and Michelson who used a maximum likelihood technique 
to estimate the wave direction \cite{Zhou_PRD_1995}.  This is not 
surprising as the two methods of estimating the wave direction have been 
shown to be equivalent (though the assumptions behind each are quite 
different) \cite{Merkowitz_PRD_1998}. The advantage of our approach is
that, by using unit vectors (Cartesian components), we are all the time
free from the anomalously high errors and correlations intrinsically
associated to the Euler angle parametrization.

Shown in Fig.~\ref{fig:solid_angle_error} is the solid angle estimation 
error as a function of the SNR.  Also shown is the results of a Monte Carlo 
type simulation of the errors.  As shown, the analytic expressions match 
the simulated errors for high SNR.  Deviations however appear for lower 
values of SNR.  This again is due to the insufficiency of the first order 
analytical estimates of the errors.  In the next section we present 
improved theoretical estimates of the errors by going one order beyond the 
first in the calculations of variances.

\section{Higher order corrections}
\label{sec:corrections}

In order to improve our theoretical understanding of the error behaviours of
the Monte Carlo simulations displayed in 
Figs.~\ref{fig:ev_error_1}-\ref{fig:solid_angle_error} we need to go one
step beyond the linear error terms of equations \eref{eq:ev_error}
and \eref{eq:deltav}. This requires calculations of higher order
derivatives for the added terms, and then the resulting general expressions
become quite complicated. They somewhat simplify for equal mode channel
variances, but are still rather cumbersome.  For example, the next order
correction to the eigenvalues, assuming the mode channel noises are
zero-mean independent Gaussian processes, is given by (see appendix)

\begin{eqnarray}
	\fl\sigma_{\lambda_k}^2 = \frac{4}{3}\,\sigma_g^2 + \nonumber\\
	+\fl\left(\sum_{m,m'=1}^5\,\frac{\partial\lambda_k}{\partial g_m}\;
	\frac{\partial^3\lambda_k}{\partial g_m\,\partial g_{m'}\,
	\partial g_{m'}} + \frac{1}{2}\,
	\frac{\partial^2\lambda_k}{\partial g_m\,\partial g_{m'}}\,
	\frac{\partial^2\lambda_k}{\partial g_m\,\partial g_{m'}}\right)\;
	\sigma_g^4,\ \ \ k=1,2,3,
	\label{eq:ev_error_2nd}
\end{eqnarray}
After rather long algebra it is found that

\begin{equation}
	\sigma_{\lambda_k}^2 = \frac{4}{3}\,\sigma_g^2 -
	2\,\left(1+\sin^2\theta_k\right)\;\frac{\sigma_g^4}{g^2}.
	\label{eq:four}
\end{equation}

It turns out that the series for $\sigma_{\lambda_k}^2$ converges {\it very
slowly\/} for low SNR, so that equation \eref{eq:ev_error_2nd} only
constitutes an improvement on (\ref{eq:ev_error}) for a rather limited range
of SNR ---see Figs.~\ref{fig:ev_error_1}-\ref{fig:ev_error_3}. Equation
\eref{eq:four} shows that the errors in $\lambda_1$ and $\lambda_2$ split
from the error in $\lambda_3$ for low values of SNR, and reproduces the
observed behaviour that $\sigma_{\lambda_3}^2$ falls below
$\sigma_{\lambda_1}^2$ and $\sigma_{\lambda_2}^2$. It is a reasonable
approximation to $\sigma_{\lambda_3}^2$ for SNR between 30 and 6, but it is
not quite as good as regards $\sigma_{\lambda_1}^2$ and $\sigma_{\lambda_2}^2$
for those values of SNR. Higher order terms would be required for an
improvement, but these imply still much longer calculations of derivatives of
the eigenvalues up to the fifth order, as can be seen in equation
(\ref{eq:sigma_f}) of the appendix.

Similar corrections can be applied to the incidence direction error estimate
of equation \eref{eq:dv3}. They appear to be given by

\begin{equation}
	|\Delta\bi{v}_3|^2 = 2\,\frac{\sigma_g^2}{g^2}\,
	\left(1+\frac{\sigma_g^2}{g^2}\right)^{-1}
	\label{eq:dve_2nd}
\end{equation}
for equal mode channel variances, $\sigma_g^2$. Solid angle errors can be
directly inferred from here:

\begin{equation}
	\Delta\Omega = \frac{2\pi}{SNR}\;\left(1+SNR^{-1}\right)^{-1}.
	\label{eq:soe_2nd}
\end{equation}

As shown in Fig.~\ref{fig:solid_angle_error}, the above theoretical
prediction is a better approach to the behaviour observed in the numerical
simulations than is equation \eref{eq:soe}. For SNR less than 2,
equation \eref{eq:soe_2nd} is also insufficient.

It is important to remind ourselves that for very low SNR the uncertainties
on the direction estimation are so large that the measurement is almost
meaningless.  Zhou and Michelson decided that a minimum SNR of 10 in energy
was necessary for a direction measurement \cite{Zhou_PRD_1995}.  Looking at
Fig.~\ref{fig:solid_angle_error} this lower limit seems reasonable, thus it
is only necessary to have accurate analytical expressions down to that level.

\section{Polarization amplitudes}
\label{sec:polarization}

We now come to the discussion of errors in the polarization amplitudes $h_+$
and $h_\times$. It is easily shown that the uncertainty on the measurement
of the polarization amplitudes is direction independent if the source
position is known ahead of time \cite{Zhou_PRD_1995}, but it has been
claimed that in the unknown direction case there is a strong direction
dependency \cite{Stevenson_PRD_1997}.  This is disturbing given that a
spherical detector is equally sensitive to waves of all polarization and
direction.  We argue in this section that the difficulties to find isotropic
estimates of $h_+$ and $h_\times$ are ultimately due to the use of unsuitable
criteria to set up those estimates. We discuss a more natural procedure,
based on data set processing, that leads to a solution of this problem.

To understand why a simple estimate of the errors leads to direction
dependencies, let us look at the basics of the solution to the inverse
problem.  We are given 3 eigenvectors and 3 eigenvalues, but these are
not independent.  The eigenvectors are orthogonal, so we actually only
get 2 pieces of information from them.  We use that information to
determine the direction of the wave.  Next, the strain tensor is traceless
so the third eigenvalue can be determined from the other two, therefore,
we only get 2 pieces of information from them.  We use one of them
(actually a combination of 2) to get the wave amplitude.  Past reasoning
suggested that we can use the last piece of information to determine the
polarization angle.  This is wrong.  The last eigenvalue does not tell us
the polarization, but rather is related to the scalar component of the GW
(or lack thereof).  We assume this to be zero for GR, so this can be
interpreted as a measurement of the non-zeroness of this component.  The
fact that we decide that this should be zero {\it a priori} does not give
us additional information about the polarization, only the level of noise in
our system.  Without any additional information we have an under-determined
system which will lead to direction dependent errors as seen in
reference~\cite{Stevenson_PRD_1997}.

The additional piece of information needed is the polarization angle
$\alpha$: the angle between the GW's axes and the eigenvectors $\bi{v}_1$
and $\bi{v}_2$ perpendicular to the incidence direction $\bi{v}_3$. To
this end, we submit our diagonal form of the matrix $\bi{A}$ to a rotation
of angle $\alpha$ about $\bi{v}_3$ to obtain best estimates of $h_+$ and
$h_\times$ by the formulas

\numparts
  \begin{eqnarray}
        h_+      & = & h\cos 2\alpha,   \label{eq:+} \\
        h_\times & = & h\sin 2\alpha,   \label{eq:x}
  \end{eqnarray}
\endnumparts
where the pure and uncorrelated noise term in $h_+$ was dropped out. The
GW amplitude $h\/$\,$\equiv$\,$(h_+^2+h_\times^2)^{1/2}$ has been shown
to have a best estimate given by equation \eref{eq:estim_h}. $h\/$ can be
determined with isotropic sensitivity since that is the case with
$\lambda_1$ and $\lambda_2$, as we have just seen. For a
{\it fixed\/}~$\alpha$, equations \eref{eq:+} and \eref{eq:x} give us an
estimate of $h_+$ and $h_\times$ in the presence of noise in the detector.

We can use the results of section \ref{sec:uncertainties_ev} to see that

\numparts
  \begin{eqnarray}
        \sigma_+^2      & = & \cos^2 2\alpha\,
                         \sigma_g^2    \label{eq:sigma+}  \\
        \sigma_\times^2 & = & \sin^2 2\alpha\,
                         \sigma_g^2    \label{eq:sigmax}  \\
        \sigma_{+\times} & = & -\frac{1}{2}\,\sin 4\alpha\,
                         \sigma_g^2    \label{eq:crossx+}
  \end{eqnarray}
\endnumparts
and these errors are indeed isotropic, for they only depend on $\sigma_g$.

In the absence of further information on the specific {\it physical\/}
nature of the source, {\it any\/} polarization angle $\alpha\/$ is valid,
for the canonical form of the tensor~(\ref{eqn:wave_strain}) is invariant
to rotations about the third axis. A particular choice of $\alpha\/$ is
thus a matter of taste in this case, and equations
\eref{eq:sigma+}--\eref{eq:crossx+} give the correct error estimates.
A common way \cite{Wagoner_Pavia_1976} to resolve the arbitrariness
in~$\alpha\/$ is to set the first Euler angle in the rotation relating
the lab frame to the wave frame equal to zero.

However, this is a very much observer dependent criterion, for detectors at
different locations would claim different values for $h_+$ and $h_\times$,
even if they agreed to be seeing the {\it same\/} source.  Errors
in $h_+$ and $h_\times$ based on such criterion have been shown e.g.\ by
\cite{Stevenson_PRD_1997} to be strongly direction dependent, which is
certainly not surprising. It is however paradoxical that a spherical detector
should prefer certain directions to others to detect a GW signal, therefore
this must be reassessed. We now propose a more consistent solution.

It is clear from the above discussion that any criterion to resolve the
arbitrariness in $\alpha\/$, therefore to estimate $h_+$ and $h_\times$,
should be established {\it relatively to the GW source\/}, be it known
ahead of time or based on a hypothesis to be checked {\it a posteriori\/}.

Let us, for concreteness, consider a {\it coalescing binary system\/} as
the GW source \cite{EugViv_PLA_1996}. The signal generated by such a system
is given by somewhat complicated functions of the space-time variables and
a number of system parameters; it will not be necessary for our purposes to
consider in detail the explicit form of such functions (see for example
\cite{Clemente_PhD_1994}), it will suffice to use formal expressions
indicating the signal dependencies:

\numparts
  \begin{eqnarray}
        h_+ & = & h_+(\bi{r},t;{\cal K}),  \\		\label{eq:cb+}
        h_\times & = & h_\times(\bi{r},t;{\cal K}).	\label{eq:cbx}
  \end{eqnarray}
\endnumparts

Here $\bi{r}\/$ is the source position, and $t\/$ is the time.
${\cal K\/}$ stands for the {\it set of characteristic source parameters\/},
which in this case include the masses of the stars, the inclination of the
orbital plane, the semimajor axis, the eccentricity of the orbit, the
periastron position, etc. Note that these amplitudes are referred to a set
of {\it source\/} axes, so they are independent of the detector's location.

The usual way to estimate the parameters ${\cal K\/}$ is to resort to
classical Statistics \cite{Helstrom_1968}, as has been done for example
in \cite{Krolak_PRD_1993} for interferometric detectors or in
\cite{Stevenson_PRD_1997} for spherical detectors. The fundamental
quantity required by such method is the {\it likelihood function\/},
$\Lambda$, which is a functional of the (unknown) signal parameters
{\it and\/} the detector data.

We then proceed as follows: we construct the likelihood function
associated to the hypothesis that equations \eref{eq:+} and \eref{eq:x}
be a fit to equations \eref{eq:cb+} and \eref{eq:cbx} for suitable values
of the parameters $\cal K\/$. It will thus have the generic form

\begin{equation}
  \Lambda = \Lambda(h;\alpha;{\cal K})    \label{eq:lambda}
\end{equation}

Standard manipulations of $\Lambda$ yield both best estimates of the signal
parameters ${\cal K\/}$ {\it and\/} of the polarization angle~$\alpha\/$,
as well as errors and cross correlations between any pair of these ---it is
recalled that such are identified as the coefficients of the {\it covariance
matrix\/}, which is the inverse of the matrix of second derivatives of
$\Lambda$ \cite{Helstrom_1968}.

We shall not attempt to give a detailed discussion of this process here.
The important point to stress is that in the approach
just proposed, we have managed to have $h\/$ as the only combination of
actual data entering the likelihood function $\Lambda$.  Errors and cross
correlations between parameter estimates will thus ultimately be functions
only of the errors and cross correlations between the eigenvalue estimates,
$\lambda_1$ and $\lambda_2$, which we have proved in
section~\ref{sec:uncertainties_ev} to be direction independent.

So not only $\alpha\/$ but also the source parameters ${\cal K\/}$ can be
determined with isotropic sensitivity by means of a spherical GW detector.
The same therefore applies to the GW amplitudes $h_+$ and $h_\times$, as
indeed expected.

The {\it quantitative\/} estimation of the errors in $h_+$ and $h_\times$
cannot however be given explicitly until the full parameter estimation
problem has been completely solved, as interactions between all those
estimates will strongly affect one another.

\section{Discussion}

With analytic expressions for the uncertainties on the eigenvectors and the
eigenvalues of the mode channel matrix~(\ref{eqn:cartesian_strain_tensor})
we can turn our attention to the physical interpretation of these values.
An unambiguous selection of $\lambda_3$ can be made on the basis that it
is the {\it third root\/} in equation \eref{eq:roots}. This will usually
be the one closest to zero. Then the other two represent the amplitude
measurement. As proved in \cite{Merkowitz_PRD_1998}, the best estimate
of the GW amplitude is the semi-difference of these two,
$(\lambda_2-\lambda_1)/2$.

The third eigenvalue ideally should be zero if general relativity is
correct.  Once noise is introduced this is no longer the case, but the
variance on this eigenvalue gives us a level of the ``non-zeroness''. One
can imagine setting a threshold on this eigenvalue that is a function
of $\sigma_{\lambda_3}$ (a function of the SNR) to veto any candidate
events that have an excessive $\lambda_3$.  Many non-GW sources are likely
to produce a non-zero $\lambda_3$, therefore becoming easily identified
and discarded.

The errors in both eigenvalues and eigenvectors are direction independent.
The last step in the analysis is the splitting of the GW amplitude $h\/$
into the usual $h_+$ and $h_\times$ components. We have shown that this
can be accomplished by making suitable reference to the source properties,
whereby isotropic sensitivity to these quantities obtains. This solves the
paradox of the anisotropies in the determination of $h_+$ and $h_\times$,
and stresses the fact that the most fundamental magnitudes to estimate
from the detector data are the eigenvalues and eigenvectors of the mode
channel matrix: these are {\it source independent\/}, and any further
progress in signal deconvolution explicitly requires reference to the
source properties, be them known ahead of time or be them stated in the
form of a hypothesis to test.

\ack{
We are grateful for the hospitality of the ROG group at the Laboratori
Nazionali di Frascati (Rome), where this work began and partly developed.
One of us (SMM) thanks E. Mauceli for helpful discussions. Two of us (JAL
and MAS) acknowledge financial support from the Spanish Ministry of
Education, contract number PB96-0384, and the Institut d'Estudis Catalans.}

\appendix

\section{Quadratic error calculations}
\label{sec:app}

Let $\bi{g}=(g_1,\,\ldots\,,g_n)$ be a set of $n\/$ independent Gaussian
random variables, with mean $\bi{\mu}=(\mu_1,\,\ldots\,,\mu_n)$ and
variances $\sigma_i^2\ (i=1,...,n)$. Let $f(\bi{g})$ be a regular
function of its arguments. Because $\bi{g}$ is a random variable so
is $f(\bi{g})$, although it will of course be generally non-Gaussian.
We want to find the mean and variance of $f(\bi{g})$, and to this end
we Taylor expand it around the mean $\bi{\mu}\/$:
\begin{equation}
    f(\bi{g}) = f(\bi{\mu}) + f_i\,\delta g_i +
      \frac{1}{2}\,f_{ij}\,\delta g_i\,\delta g_j +
      \frac{1}{6}\,f_{ijk}\,\delta g_i\,\delta g_j\,\delta g_k
      + \ldots
      \label{eq:taylor}
\end{equation}
where
\begin{equation}
    f_{ij\ldots}\equiv\left.\frac{\partial f}
    {\partial g_i\partial g_j\ldots}\right|_{\bi{g}=\bi{\mu}}\ \ \ 
    \mbox{and}\ \ \ \delta g_i\equiv g_i - \mu_i.
    \label{eq:defin}
\end{equation}
and the usual convention of summation over repeated indices is adopted
in~(\ref{eq:taylor}).

The mean of $f(\bi{g})$ is its expectation value, $E\{f(\bi{g})\}$,
while its variance is the difference
\begin{equation}
    \sigma_f^2 = E\{f^2(\bi{g})\} - [E\{f(\bi{g})\}]^2
    \label{eq:ms}
\end{equation}

Expectation values are to be taken on the basis of the
expansion~(\ref{eq:taylor}). Given that $\bi{g}$ is a set of independent
Gaussian variables, the expectation of the product of an {\it odd\/} number
of $\delta g$'s is zero, while
\begin{eqnarray}
    E\{\delta g_i\,\delta g_j\} & = & \delta_{ij}\,\sigma_i^2 \\
    E\{\delta g_i\,\delta g_j\,\delta g_k\,\delta g_l\} & = &
    \delta_{ij}\,\delta_{kl}\;\sigma_i^2\,\sigma_k^2 +
    \delta_{ik}\,\delta_{jl}\;\sigma_i^2\,\sigma_j^2 +
    \delta_{il}\,\delta_{jk}\;\sigma_i^2\,\sigma_j^2
    \label{eq:delta}
\end{eqnarray}
etc., where no summation over repeated indices is exceptionally assumed
in these expressions. If the assumption is made that all the $g\/$'s
have equal variances, $\sigma^2$, then one easily finds
\begin{equation}
    E\{f(\bi{g})\} = f + \frac{1}{2}\,f_{ii}\;\sigma^2
      + \frac{1}{8}\,f_{iijj}\;\sigma^4
      + \frac{1}{48}\,f_{iijjkk}\;\sigma^6 + \ldots
    \label{eq:meanf}
\end{equation}
where $f\/$ is a shorthand for $f(\bi{\mu})$. Likewise,

\begin{eqnarray}
    \fl\left[E\{f(\bi{g})\}\right]^2  =  f^2 + ff_{ii}\;\sigma^2 +
      \frac{1}{4}\,\left(ff_{iijj}+f_{ij}f_{ij}\right)\,\sigma^4 \nonumber \\
      +\frac{1}{24}\,\left(ff_{iijjkk}+3\,f_{ij}f_{ijkk}\right)\,\sigma^6
      + \ldots,
\end{eqnarray}
and

\begin{eqnarray}
    \fl E\{f^2(\bi{g})\}  =  f^2 + \left(ff_{ii}+f_if_i\right)\,\sigma^2 +
      \frac{1}{4}\,\left(ff_{iijj}+4\,f_if_{ijj}+3\,f_{ij}f_{ij}\right)
      \,\sigma^4 \nonumber \\
      +\frac{1}{24}\,\left(ff_{iijjkk}+6\,f_if_{ijjkk}+15\,f_{ij}f_{ijkk}+
          10\,f_{ijk}f_{ijk}\right)\,\sigma^6 + \ldots
\end{eqnarray}
So, finally,

\begin{eqnarray}
    \fl\sigma_f^2 = f_if_i\;\sigma^2 +
      \left(f_if_{ijj}+\frac{1}{2}\,f_{ij}f_{ij}\right)\,\sigma^4bb
      \nonumber \\
      \lo+\frac{1}{4}\,\left(f_if_{ijjkk}+2\,f_{ij}f_{ijkk}+
      \frac{5}{3}\,f_{ijk}f_{ijk}\right)\,\sigma^6 + \ldots
    \label{eq:sigma_f}
\end{eqnarray}

\end{document}